# Optical aperture synthesis with electronically connected telescopes

Dainis Dravins[1], Tiphaine Lagadec[1,*] & Paul D. Nuñez[2,3,§]

[1] Lund Observatory, Lund University, Box 43, SE-22100 Lund, Sweden
[2] Collège de France, 11 Place Marcelin Berthelot, FR-75005 Paris, France
[3] Laboratoire Lagrange, Observatoire de la Côte d'Azur, BP 4229, FR-06304 Nice, France
[*] Present address: ESTEC, European Space Research and Technology Centre, Keplerlaan 1, NL-2200 AG Noordwijk, The Netherlands
[§] Present address: JPL, Jet Propulsion Laboratory, California Institute of Technology, 4800 Oak Grove Drive, Pasadena, CA 91109-8099, U.S.A.

**Abstract**

Highest resolution imaging in astronomy is achieved by interferometry, connecting telescopes over increasingly longer distances, and at successively shorter wavelengths. Here, we present the first diffraction-limited images in visual light, produced by an array of independent optical telescopes, connected electronically only, with no optical links between them. With an array of small telescopes, second-order optical coherence of the sources is measured through intensity interferometry over 180 baselines between pairs of telescopes, and two-dimensional images reconstructed. The technique aims at diffraction-limited optical aperture synthesis over kilometre-long baselines to reach resolutions showing details on stellar surfaces and perhaps even the silhouettes of transiting exoplanets. Intensity interferometry circumvents problems of atmospheric turbulence that constrain ordinary interferometry. Since the electronic signal can be copied, many baselines can be built up between dispersed telescopes, and over long distances. Using arrays of air Cherenkov telescopes, this should enable the optical equivalent of interferometric arrays currently operating at radio wavelengths.

**Introduction**

In optical astronomy, the highest angular resolution is presently offered by phase/amplitude interferometers combining light from telescopes separated by baselines up to a few hundred metres. Tantalizing results show how stellar disks start to become resolved, revealing stars as a diversity of individual objects, although until now feasible only for a small number of the largest ones. Concepts have been proposed to extend such facilities to scales of a km or more, as required for surface imaging of bright stars with typical sizes of a few milliarcseconds. However, their realization remains challenging, both due to required optical and atmospheric stabilities within a fraction of an optical wavelength, and the need to span many interferometric baselines (given that optical light cannot be copied with retained phase, but must be divided and diluted by beamsplitters to achieve interference among multiple telescope pairs). While atmospheric issues could be avoided by telescope arrays in space, such are impeded by their complexity and cost. However, atmospheric issues can be circumvented by measuring higher-order coherence of light through intensity interferometry.

In the following, we present a laboratory demonstration of a multi-telescope array to verify the end-to-end sequence of operation from observing star-like sources, to reconstruction of their images.

## Origin and principles of intensity interferometry

The technique of intensity interferometry[1], once pioneered by Hanbury Brown and Twiss is effectively insensitive to atmospheric turbulence and permits very long baselines at also short optical wavelengths. What is observed is the second-order coherence of light (that of intensity *I*, not of amplitude nor phase), by computing temporal correlations of arrival times between photons recorded in different telescopes observing the same target. When the telescopes are close together, the intensity fluctuations measured in both telescopes are correlated in time, but when further apart, the degree of correlation diminishes.

Although this behaviour can be understood in terms of classical optical waves undergoing random phase shifts, it is fundamentally a two-photon process measuring the degree of photon 'bunching' in time, and its full explanation requires a quantum description. The concept is normally seen as the first experiment in quantum optics[2-4], with its first realization already in the 1950's, when Hanbury Brown and Twiss succeeded in measuring the diameter of the bright and hot star Sirius[5]. This laid the foundation for experiments of photon correlations including also states of light that do not have classical counterparts (such as photon antibunching), and has found numerous applications outside optics, in particular in particle physics since the same bunching and correlation properties apply to other bosons, i.e., particles with integer quantum spin, whose statistics in an equilibrium maximum-entropy state follow the Bose-Einstein distribution. The method can also be seen as mainly exploiting the corpuscular rather than the wave nature of light, even if those properties cannot be strictly separated.

The quantity measured is $<I_1(t) \bullet I_2(t)> = <I_1(t)><I_2(t)> (1 + |\gamma_{12}|^2)$, where $<>$ denotes temporal averaging and $\gamma_{12}$ is the mutual coherence function of light between locations 1 and 2, the quantity commonly measured in phase/amplitude interferometers. Compared to randomly fluctuating intensities, the correlation between intensities $I_1$ and $I_2$ is 'enhanced' by the coherence parameter, and an intensity interferometer thus measures $|\gamma_{12}|^2$ with a certain electronic time resolution. Realistic values of a few nanoseconds correspond to light-travel distances around one metre, and permitted error budgets relate to such a number rather than to the optical wavelength, making the method practically immune to atmospheric turbulence and optical imperfections. The above coherence relation holds for ordinary thermal ('chaotic') light where the light wave undergoes 'random' phase jumps on timescales of its coherence time but not necessarily for light with different quantum statistics[2-4].

## Intensity interferometry in the modern era

The method has not been pursued in astronomy since Hanbury Brown's early two-telescope intensity interferometer in Narrabri, Australia[1], but new possibilities are opening up with the erection of large arrays of air Cherenkov telescopes. Although primarily devoted to measuring optical Cherenkov light in air induced by cosmic gamma rays, their optical performance, telescope sizes, and distributions on the ground, would be suitable for also such applications[6].

Realizing that potential, various theoretical simulations of intensity interferometry observations with extended telescope arrays have been made[7-11], in particular examining the overall sensitivity and limiting stellar magnitudes. Different studies conclude that, with current electronic performance and current Cherenkov telescope designs, one night of observing in one single wavelength band permits measurements of hotter stars down to visual magnitude about $m_v = 8$, giving access to thousands of sources. Such values agree with extrapolations from the early pioneering observations, also when including observational issues such as background light from the night sky[10]. Fainter sources could be reached once simultaneous measurements in multiple wavelength bands can be realized, or if full two-dimensional data are not required.

**Interferometer array in the laboratory**

In this work, we report the first end-to-end practical experiments of such types of measurements. A setup was prepared in a large optics laboratory with artificial sources ('stars'), observed by an array of small telescopes of 25 mm aperture, each equipped with a photon-counting solid-state detector (silicon avalanche photodiode operated in Geiger-mode). The continuous streams of photon-counts were fed into electronic firmware units in real time (with 5 ns time resolution) computing temporal cross correlations between the intensity fluctuations measured across baselines between many different pairs of telescopes. The degree of mutual correlation for any given baseline provides a measure of the second-order spatial coherence of the source at the corresponding spatial frequency. Numerous telescope pairs of different baseline lengths and orientations populate the interferometric Fourier-transform plane, and the measured data provide a two-dimensional map of the second-order spatial coherence of the source, from which its image can be reconstructed.

**Experiment setup**

Artificial 'stars' (single and double, round and elliptic) were prepared as tiny physical holes drilled in metal (using miniature drills otherwise intended for mechanical watches). Typical aperture sizes of 0.1 mm subtend an angle of ~1 arcsec as observed from the 23 metre source-to-telescope distance in the laboratory.

The technique of intensity interferometry, in principle, requires chaotic light for its operation[2-4] in order for 'random' intensity fluctuations to occur and, in practice, also sources of high surface brightness (such as stars hotter than the Sun). This brightness requirement of a high effective source temperature comes from the need for telescopes to be small enough to resolve the structures in the Fourier domain, while still receiving sufficiently high photon fluxes. Of course, the same limitations apply to laboratory sources as to celestial ones. There are numerous hot stars in the sky but correspondingly hot laboratory sources are less easy to find. Ordinary sources producing chaotic light have too low effective temperatures (e.g., high-pressure arc lamps of around 5,000 K) to permit conveniently short integration times while laser light cannot be directly used since it is not chaotic (ideally, undergoing no intensity fluctuations whatsoever).

Following various tests, the required brilliant illumination was produced by scattering $\lambda = 532$ nm light from a 300 mW laser against microscopic particles in thermal (Brownian) motion. Those were monodispersive polystyrene spheres of 0.2 μm diameter, suspended in room-temperature distilled water, undergoing random motion due to collisions with the water molecules surrounding them. Upon scattering, the laser light is broadened through Doppler shifts, producing a spectral line with a Lorentzian wavelength shape, and with a chaotic (Gaussian) distribution of the electric field amplitudes[12,13]. Such light is thus equivalent in its photon statistics and intensity fluctuations to the thermal (white) light expected from any normal star, the difference being that the spectral passband is now very much narrower. The width of the scattered line was estimated from its measured temporal coherence to around 10-100 kHz, with a brightness temperature on order $10^5$ K. The relatively long coherence times enabled convenient measurements on microsecond scales, with typical photon-count rates between 0.1 – 1 MHz, integrating during a few minutes for each telescopic baseline. Such count rates assure negligible photon-noise contributions although there may still remain issues of systematics due to, e.g., slightly inhomogeneous illumination of the artificial stars.

Here we note that the signal-to-noise ratio in intensity interferometry in principle is independent of optical bandpass, be it broadband white light or just a narrow spectral line[1]. For any broader bandpass, realistic electronic time resolutions are slower than optical coherence times, and the noisier measurement of a smaller photon flux from a narrower passband (with longer coherence time) is compensated by lesser averaging over fewer coherence times. Thus, also the present quasi-monochromatic light source can represent measurements over broader optical bandpasses.

**Simulating a large telescope array**

Angular resolution in optical systems is often quantified by the Airy disk diffraction radius $\theta \approx 1.22\, \lambda/D$ rad. At $\lambda = 532$ nm, the baseline $D$ required to resolve 1 arcsec is $\sim 10$ cm, dictating a rather compact setup of small telescopes. Our five units offered baselines between 3 and 20.5 cm, the former constraining the largest field to nominally 4.5 arcsec, and the latter limiting the resolution to 0.65 arcsec. The zero baseline (needed for calibration) was realized with two additional separate telescopes behind one beamsplitter, both telescopes observing the same signal.

To mimic a large telescope array, the effective number of telescopes and baselines was increased by measuring across different interferometer angles. The array was mounted on a horizontal optical table and thus covered horizontal baselines only. To cover also oblique and vertical baselines in the two-dimensional Fourier-transform plane, the position angle of the artificial 'star' was successively rotated relative to the plane of the telescopes. Various details of the overall laboratory arrangements and of measuring techniques are described elsewhere[14]. With $N$ telescopes, $N(N-1)/2$ different baselines can be constructed. Five telescopes give 10 baselines at any given angle, and a series of 18 measurement in steps of 10 degrees over a total of 180° produces 180 baselines. The second-order coherence in a point $(u,v)$ of the Fourier-transform plane equals that in $(-u,-v)$, so each baseline provides two points; data for angles up to 360° are copies of the first ones. Figure 1 shows the coherence pattern for one thus measured asymmetric binary 'star', with one larger and one smaller component (flux ratio 4:1). No smoothing was applied to the measured coherence values, neither for this plot, nor for the later image reconstructions. The overplotted surface merely visualizes the global measured pattern, whose individual points are too numerous to be readily displayed.

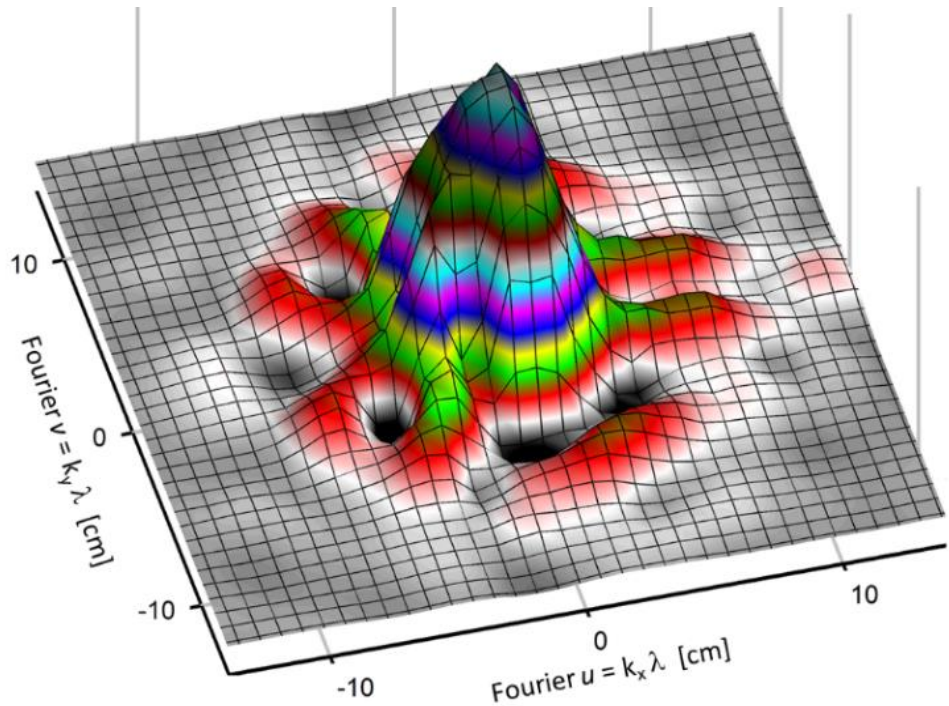

**Figure 1** Second-order optical coherence surface Measurements for an artificial asymmetric binary star built up from intensity-correlation measurements over 180 baselines between pairs of small optical telescopes in the laboratory. The coordinates refer to the plane of the Fourier transform of the source image and correspond to different telescope baseline lengths and orientations. The image reconstructed from these data is in Figure 2, right

**Image reconstruction principles**

Image reconstruction from second-order coherence implies some challenges not present in first-order phase/amplitude techniques. While intensity interferometry has the advantage of not being sensitive to phase errors in the optical light paths, it also does not measure such phases but rather obtains the absolute magnitudes of the respective Fourier-transform components of the source image. Such data can by themselves fit model parameters such as stellar diameters or binary separations but images cannot be directly computed through merely an inverse Fourier transform.

Theoretical simulations have verified that, provided the Fourier-transform plane is well filled with data, two-dimensional image restoration becomes feasible[8,9,11]. The pattern of second-order coherence is equivalent to the intensity pattern produced by diffraction of light in a correspondingly sized aperture, and it is already intuitively clear that a thorough mapping of that pattern must put stringent constraints on the source geometry.

**Image reconstruction procedures and results**

Strategies in analyzing intensity interferometry data involve estimating Fourier phases from Fourier magnitudes[8,15,16]. Since the Fourier transform of a finite object is an analytic function, the Cauchy-Riemann equations can be used to find derivatives of the phase from derivatives of the magnitude, producing images that are unique except for translation and reflection. Here, one could start with a one-dimensional phase estimate along a single slice through the ($u$,$v$)- plane origin. 2-D coverage requires combining multiple 1-D reconstructions, while ensuring mutual consistency between adjacent ones. However, these algorithms appear to require a very dense coverage of the ($u,v$)-plane and, despite our numerous baselines, did not appear efficient with current data.

Instead, an inverse-problem approach was taken, imposing very general constraints to interpolate between Fourier frequencies ('regularization'). The Multi-aperture Image Reconstruction Algorithm, MiRA (previously tested on simulated data[9]), produced more stable results. This iterative procedure maximizes the agreement between measurements and the squared modulus of the Fourier transform of the reconstructed image. Regularization techniques can be applied to favour solutions with certain properties, e.g., if something is initially known about the source. To reconstruct the elliptical star in Figure 2, a 'compact' solution (quadratic regularization) was favoured, and a 'smooth' solution for the binary (quadratic smoothness regularization)[17].

Figure 2 shows examples of such image reconstructions at λ = 532 nm from measurements with 100 and 180 optical baselines. As far as we are aware, these are the first diffraction-limited images in visual light that have been reconstructed from an array of dispersed optical telescopes, connected by electronic software only, without any optical links between them.

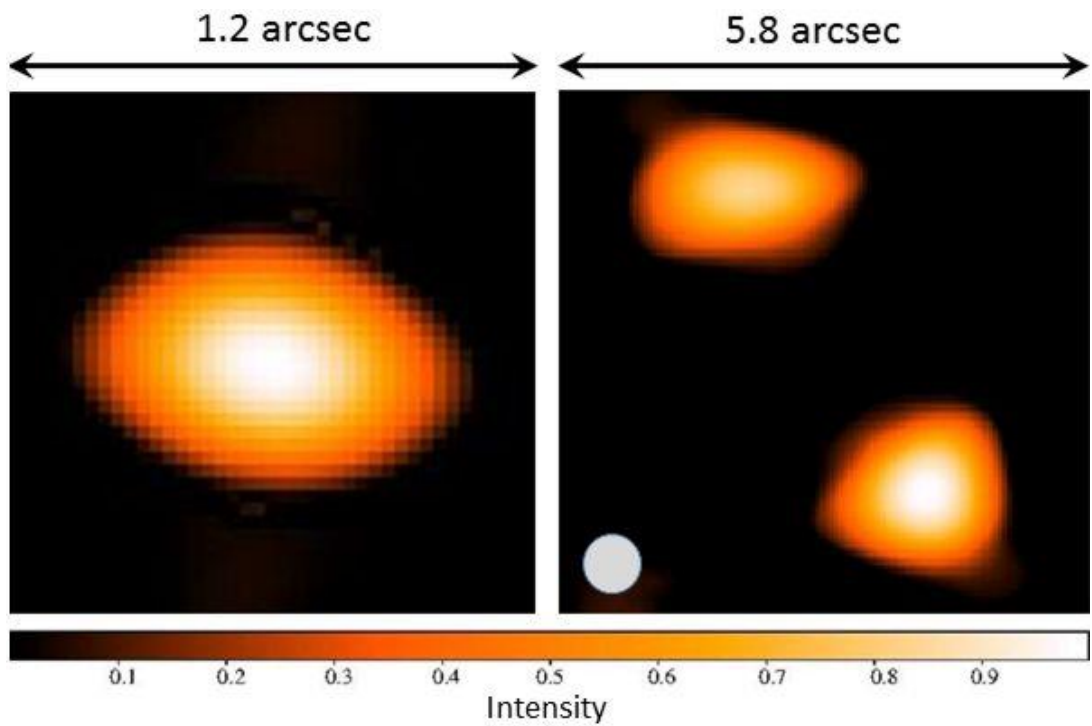

**Figure 2** Optical images reconstructed from electronic intensity interferometry Measurements with 100 and 180 baselines, respectively, of a single elliptical artificial 'star', and an asymmetric binary one (flux ratio 4:1). Sufficiently dense coverage of the interferometric Fourier-transform plane enables a full reconstruction of source images, despite the lack of Fourier phase information. In principle, the images are unique except for their possible mirrored reflections. The spatial scales are similar on both axes; colours show reconstructed linear intensities, the circle indicates the 0.65 arcsec resolution set by the longest baseline.

Similar to any phase/amplitude interferometry or aperture synthesis, the fidelity of the reconstructed images depends on measurement noise, the density of coverage across the Fourier ($u$,$v$)-plane, and the character of the 'cleaning' algorithms applied. In the present measurements, the random part of the noise is essentially negligible but some systematics are likely present, e.g., due to a probably somewhat inhomogeneous illumination of the artificial stars that makes their brightness shapes not identical to their geometrical apertures.

Reconstructions were tested on different sources. For the single elliptic star (Figure 2, left), its major and minor axes could be determined to a precision of 2%, a number estimated from several image reconstructions with different initial random images. Even slightly better values were obtained for a symmetric binary. However, the present asymmetric one was chosen to be a more difficult target, because of the contrast between its components, and since the Fourier-plane coverage is marginal for both the largest and smallest scales, contributing to some image artefacts seen in Figure 2. Still, the reconstruction precision is consistent with what is found in previous detailed theoretical simulations[8], where the reconstructed radii gradually become less precise for greater brightness ratios between both binary members.

The computing effort is modest (and not comparable to the extensive numerical efforts involved in radio aperture synthesis that involve also spectral analysis). To obtain an image of 10,000 pixels from measurements across 100 baselines, starting from a random image, requires just some 10 seconds on a 3 GHz processor. The exact time can be shorter if one already has some initial guess (perhaps obtained from some 'raw' reconstruction from a Cauchy-Riemann analysis), or be longer for additional baselines or more numerous pixels.

The image quality was further explored as function of the sampling or smoothing in the ($u$,$v$)-plane, affecting the pixel sizes in the reconstructed image. The best performance was found when the number of samples within the Fourier plane roughly matches the number of baseline measurements. Possibly, there could be other dependences for noise-limited data, where some kind of smoothing may be required. In general, the optimization of reconstruction for different numbers of sample points in the presence of various noise levels and incomplete Fourier-plane coverages is a somewhat complex problem that will require further studies (analogous to what in the past has been made for aperture synthesis in the radio).

**Long-baseline optical interferometry**

The longer-term ambition is to connect widely separated telescopes in a large optical aperture synthesis array. Similar to current radio interferometry, the electronic signal from any telescope can be freely copied or even stored for possible later analysis. Realistic signal bandwidths for measured intensity fluctuations are on the order of 100 MHz, very comparable to those encountered in radio arrays. However, the requirements on the signal correlators are much more modest than for radio phase/amplitude interferometers. Optical intensity interferometry can in practice only detect spatial coherence while the wavelengths are selected by colour filters or through detector responses, not – as the case in radio – by temporal coherence measurements. Consequently, the correlator electronics can be simple table-top devices rather than massive computing facilities.

The signal-to-noise properties[1,6,7] of intensity interferometry require the telescopes to be large in order to reach adequately low photon noise, but their optical precision needs not be higher than that of ordinary air Cherenkov telescopes. However, the limiting signal-to-noise ratios may still depend on their exact optical design. For example, how high electronic time resolutions that can be exploited may be constrained by the time-spread of light-paths inside the telescope. Details of such and other parameters are discussed elsewhere[6,7,10].

An obvious candidate for a kilometre-sized optical imager are the telescopes of CTA, the planned Cherenkov Telescope Array[18], foreseen to have on the order of 100 telescopes, distributed over a few square km. Assuming it becomes available for also interferometric observations, the spatial resolution at the shortest optical wavelengths may approach ~ 30 µas. Such resolutions have hitherto been reached only in the radio, and it is awkward to predict what features on, or around stellar surfaces will appear in the optical[6]. However, to appreciate the meaning of such resolutions, we note that a hypothetical Jupiter-size exoplanet in transit across some nearby bright star such as Sirius would subtend an easily resolvable angle of some 350 µas[14]. While spatially resolving the disk of an exoplanet in its reflected light may remain unrealistic for the time being, the imaging of its dark silhouette on a stellar disk – while certainly very challenging[19] – could perhaps be not quite impossible.

**References**

**1.** Hanbury Brown, R., *The Intensity Interferometer. Its Applications to Astronomy.* Taylor & Francis, London (1974)

**2.** Labeyrie, A., Lipson, S. G. & Nisenson, P., *An Introduction to Optical Stellar Interferometry.* Cambridge University Press, Cambridge (2006)

**3.** Saha, S. K., *Aperture Synthesis. Methods and Applications to Optical Astronomy.* Springer, New York (2011)

**4.** Shih, Y., *An Introduction to Quantum Optics. Photon and Biphoton Physics.* CRC Press, Boca Raton (2011)

**5.** Hanbury Brown, R. & Twiss, R. Q., A test of a new type of stellar interferometer on Sirius. Nature, **178**, 1046-1048 (1956)

**6.** Dravins, D., LeBohec, S., Jensen, H. & Nuñez, P. D., for the CTA Consortium, Optical intensity interferometry with the Cherenkov Telescope Array. Astropart. Phys. **43**, 331-347 (2013)

**7.** Dravins, D., LeBohec, S., Jensen, H. & Nuñez, P. D., Stellar intensity interferometry: Prospects for sub-milliarcsecond optical imaging. New Astron. Rev. **56**, 143-167 (2012)

**8.** Nuñez, P. D., Holmes, R., Kieda, D. & LeBohec, S., High angular resolution imaging with stellar intensity interferometry using air Cherenkov telescope arrays. Mon. Not. Roy. Astron. Soc. **419**, 172-183 (2012)

**9.** Nuñez, P. D., Holmes, R., Kieda, D., Rou, J. & LeBohec, S., Imaging submilliarcsecond stellar features with intensity interferometry using air Cherenkov telescope arrays. Mon. Not. Roy. Astron. Soc. **424**, 1006-1011 (2012)

**10.** Rou, J., Nuñez, P. D., Kieda, D. & LeBohec, S., Monte Carlo simulation of stellar intensity interferometry. Mon. Not. Roy. Astron. Soc. **430**, 3187-3195 (2013)

**11.** Dolne, J.J., Gerwe, D. R. & Crabtree, P.N., Cramer-Rao lower bound and object reconstruction performance evaluation for intensity interferometry. Proc. SPIE **9146**, Optical and Infrared Interferometry IV, 914636-1 – 914636-12 (2014)

**12.** Berne, B. J. & Pecora, R., *Dynamic Light Scattering. With Applications to Chemistry, Biology and Physics.* Dover, Mineola, NY (2000)

**13.** Crosignani, B., Di Porto, P. & Bertolotti, M., *Statistical Properties of Scattered Light.* Academic Press, New York (1975)

**14.** Dravins, D. & Lagadec, T., Stellar intensity interferometry over kilometer baselines: Laboratory simulation of observations with the Cherenkov Telescope Array. Proc. SPIE **9146**, Optical and Infrared Interferometry IV, 91460Z-1 – 91460Z-18 (2014)

**15.** Holmes, R. B. & Belen'kii, M. S., Investigation of the Cauchy-Riemann equations for one-dimensional image recovery in intensity interferometry. J. Opt. Soc. Am. A **21**, 697-706 (2004)

**16.** Holmes, R., Calef, B., Gerwe, D. & Crabtree, P., Cramer-Rao bounds for intensity interferometry measurements. Appl. Opt. **52**, 5235-5246 (2013)

**17.** Thiébaut, E., Image reconstruction with optical interferometers. New Astron. Rev. **53**, 312-328 (2009)

**18.** Actis, M. et al., Design concepts for the Cherenkov Telescope Array CTA: An advanced facility for ground-based high-energy gamma-ray astronomy. Exp. Astron. **32**, 193-316 (2011)

**19.** Strekalov, D. V., Erkmen, B. I. & Yu, N., Intensity interferometry for observation of dark objects. Phys. Rev. A **88**, id. 053837, 9 pp. (2013)

**Acknowledgements**

This work was supported by the Swedish Research Council and The Royal Physiographic Society in Lund. Early experiments in intensity interferometry and photon correlation techniques at Lund Observatory involved also Toktam Calvén Aghajani, Daniel Faria, Johan Ingjald, Hannes Jensen, Lennart Lindegren, Eva Mezey, Ricky Nilsson and Helena Uthas.

**Author Contributions**

D.D. conceived this project; D.D. and T.L. together set up the laboratory experiment at Lund Observatory, performed all measurements and their reductions; P.D.N. carried out the image reconstructions. D.D. wrote the draft manuscript but all authors contributed to editing its final version.

**Author Information**

Reprints and permissions information is available at www.nature.com/reprints . The authors declare no competing financial interests. Correspondence and requests for materials should be addressed to dainis@astro.lu.se .